\documentclass[manuscript]{geophysics}
\usepackage{mathrsfs,amsmath, multirow, xspace}
\usepackage{hyperref}
\usepackage{makecell}
\usepackage{lineno}

\newcommand{\SimPEG}{\textsc{SimPEG}\xspace}

\newcommand{\curl}{{\vec \nabla}\times}
\newcommand {\J}{{\vec J}}

\newcommand {\E}{{\vec E}}
\newcommand{\siginf}{\sigma_\infty}
\newcommand{\dsig}{\triangle\sigma}
\newcommand{\dcurl}{{\mathbf C}}

\newcommand{\M}{{\mathbf M}}
\newcommand{\MfMui}{{\M^f_{\mu^{-1}}}}

\newcommand{\Me}{{\M^e}}

\newcommand{\Mes}[1]{{\M^e_{#1}}}

\renewcommand {\j}  { {\vec j} }

\renewcommand {\b}  { {\vec b} }
\newcommand {\e}  { {\vec e} }

\renewcommand {\dj}  { {\mathbf{j} } }

\newcommand {\db}  { {\mathbf{b} } }
\newcommand {\de}  { {\mathbf{e} } }

\newcommand{\dip}{d^{IP}}

\begin{document}

\title{Detecting induced polarization effects in time-domain data: a modeling study using stretched exponentials}

\renewcommand{\thefootnote}{\fnsymbol{footnote}}

\ms{...} 

\address{
\footnotemark[1]Geophyscial Inversion Facility, \\
University of British Columbia, Canada\\
BC, Vancouver}

\author{Seogi Kang \footnotemark[1],\footnotemark[2], Douglas W. Oldenburg\footnotemark[2] and Lindsey J. Heagy\footnotemark[2]\\[16pt]
{\normalfont \small
\footnotemark[1]
Corresponding author: skang@eos.ubc.ca \\
\footnotemark[2]
Geophysical Inversion Facility, University of British Columbia
}\\[16pt]
{
\normalfont \small
}
}

\footer{Detecting induced polarization effects in AEM data}
\lefthead{Kang et al.}
\righthead{Detecting induced polarization effects in AEM data}

\maketitle
\begin{abstract}
The potential for extracting and interpreting induced polarization (IP) data from airborne surveys is now broadly recognized. There is, however, still considerable discussion about the conditions under which the technique can provide knowledge about the subsurface and thus, its practical applications. Foremost among these is whether, or under what conditions, airborne IP can detect chargeable bodies at depth. To investigate, we focus on data obtained from a coincident-loop time-domain system. Our analysis is expedited by using a stretched exponential rather than a Cole-Cole model to represent the IP phenomenon. Our paper begins with an example that illuminates the physical understanding about how negative transients (the typical signature of an IP signal in airborne data) can be generated. The effects of the background conductivity are investigated; this study shows that a moderately conductive and chargeable target in a resistive host is an ideal scenario for generating strong IP signals. We then examine the important topic of estimating the maximum depth of the chargeable target that can generate negative transients. Lastly, some common chargeable earth-materials are discussed and their typical IP time-domain features are analyzed. The results presented in this paper can be reproduced and further explored by accessing the provided Jupyter notebooks.
\end{abstract}

\renewcommand{\figdir}{./figures} 


\section{Introduction}
Some earth  materials  are chargeable because they can store charge when an electric field is applied by an electromagnetic (EM) source. This is often called the induced polarization (IP) phenomenon. These materials can have different polarization mechanisms which results in different IP characteristics as a function of frequency. This can be translated into a complex conductivity model such as the Cole-Cole conductivity model \cite[]{cole1941,pelton1978,tarasov2013}:
\begin{equation}
    \sigma_{cc}(\omega) = \siginf - \frac{\eta_{cc} \siginf}{1+(\imath \omega \tau_{cc})^{c_{cc}}}
    \label{eq:colecole}
\end{equation}
where $\siginf$ is the conductivity at infinite frequency,  $\eta_{cc}$ is the chargeability, $\tau_{cc}$ is the time constant (s), and  $c_{cc}$ is the frequency exponent; the subscript CC indicates Cole-Cole.

IP surveys have been successfully conducted in a variety of geoscience applications. For mining, IP surveys are recognized as a principal geophysical technique for finding disseminated sulphides or porphyry deposits \cite[]{fink1990}. Non-metallic materials such as clays and ice can also generate IP signals \cite[]{grimm2015,leroy2009}; this makes IP a useful technique in a range of environmental applications. Grounded DC-IP surveys have been successfully used for both mining and environmental applications for the past decades.
Airborne EM (AEM) systems can also detect IP signals.
In particular time-domain AEM surveys using a coincident-loop system sometimes display a negative transient; this is a distinctive IP signature \cite[]{weidelt1982}. Compared to EM signals, these negatives (IP signals) are much smaller in amplitude. Hence, for the initial AEM systems, it was not clear if the measured negatives were signals from chargeable materials or if they were simply noise generated by power lines or electric fences \cite[]{smith1996}. With time however, instruments have improved and the validity of negative transients as signal has been firmly established \cite[]{macnae2016,viezzoli2017}. For instance, consistent negatives were recorded over the Tli Kwi Cho kimberlite deposit with three different AEM systems \cite[]{kang2017}. As the quality of instrumentation improves, it is expected that more IP signals will be measured in airborne data. This ability provides motivation for developing methodologies that can extract chargeability information from airborne IP data. Various approaches, including simple curve-fitting, 1D inversions, and 3D inversions have been developed and successfully applied to field examples \cite[]{kratzer2012,kwan2015,hodges2014,kaminski2017,kang2017b}. There is a significant enthusiasm for the potential use of the airborne IP techniques in a variety of applications (e.g. mining and groundwater). However, setting proper expectations about the technique, and recognizing its limitation based upon the physics and the current system specifications, is crucial because neither overselling nor underselling the technique is beneficial for the community.

\cite{macnae2016} investigated the physics of airborne IP and its practical aspects using a simple thin-sheet solution. A main conclusion from his study was that airborne IP is effectively a surficial clay mapper ($z<$ 100 m). \cite{viezzoli2017}, showed the potential that a deeper chargeable target, such as a mineral deposit ($z>$ 100 m), can be detected. That work however, was based upon analysis using 1D simulations. Hence, there is disagreement about the potential depth of investigation of the airborne IP technique. Although the approximate thin-sheet solution and semi-analytic 1D solutions used in \cite{macnae2016} and \cite{viezzoli2017}, respectively, can illustrate some meaningful concepts with respect to airborne IP, these approaches are limited in their ability to model the physics in the presence of complex conductivity structures. For instance, the finite size of the chargeable structure (e.g. width and length) is not taken into account in either approach. Investigating the feasibility of airborne IP in realistic geologic settings requires the use of 3D numerical simulations that solve the full Maxwell's equations.

In this paper, we first develop a convolutional time-domain EM (TEM) simulation code using a stretched exponential (SE) conductivity function \cite[]{kohlrausch1854}. We then use this code to investigate four main questions related to the feasibility of the airborne IP under ranges of circumstances:
\begin{itemize}
  \item How does chargeable material in the subsurface generate negative transients in coincident loop systems?
  \item How does the background conductivity affect the IP signals?
  \item To what depth can we expect to detect a chargeable body?
  \item What are the characteristics of detectable chargeable materials in AEM data?
\end{itemize}
For our feasibility study, we limit our attention to detectability of IP signals, and we do not address issues of resolvability of chargeable structures in the inversion; that  issue is beyond the scope of this study.

\section{SIMULATING AIRBORNE IP DATA USING A STRETCHED EXPONENTIAL}
With a complex conductivity, $\sigma(\omega)$, the current density, $\J$, in the frequency domain, can be written as:
\begin{equation}
    \J = \sigma(\omega)\E
    \label{eq:ohms_law_frequency}
\end{equation}
where $\E$ is the electric field (V/m). In the time-domain, the current density, $\j$, is:
\begin{equation}
    \j = \sigma(t) \otimes \e
    \label{eq:ohms_law_time}
\end{equation}
where $\otimes$ is a convolution. Then Maxwell's equations can be written as
\begin{equation}
    \curl \e = -\frac{\partial \b}{\partial t}
    \label{eq:faraday}
\end{equation}
\begin{equation}
    \curl \mu^{-1}\b - \j = \j_s
    \label{eq:ampere}
\end{equation}
where $\b$ is the magnetic flux density (Wb/m$^2$) and $\j_s$ (A/m$^2$) is the current source;  $\mu$ is the magnetic permeability (H/m). By discretizing and solving the above equations in 3D, we can compute TEM data that include IP effects \cite[]{marchant2014,marchant2015}. For the discretization of eqs. (\ref{eq:ohms_law_frequency})-(\ref{eq:ampere}), an open-source geophysical simulation and inversion package, SimPEG, is used \cite[]{cockett2015}. The developed \textsc{SimPEG-EMIP} code works for both 3D tensor meshes and 2D/3D cylindrical meshes \cite[]{heagy2017}. Although not shown in the paper, the  \textsc{SimPEG-EMIP} code can handle arbitrary waveforms such that user can input actual system waveforms used for their own case studies. For further details about solving the convolutional form of Maxwell's equations, see Appendix A. The code is tested with an analytic solution described in Appendix B.

For a time-dependent conductivity, $\sigma(t)$, we use the stretched exponential (SE) model rather than the Cole-Cole model defined in the frequency-domain (eq. \ref{eq:colecole}). The SE conductivity for a step-off function, $1-u_{step}(t)$, can be written as
\begin{equation}
    \sigma_{se} \otimes (1-u_{step}) =
    \sigma_0 \Big(1-u_{step}(t)\Big)
    -\siginf \eta_{se}\rm{exp}\Big(-\left(\frac{t}{\tau_{se}}\right)^{c_{se}}\Big)u_{step}(t)
    \label{eq: sigma_se_step_off}
\end{equation}
where $u_{step}(t)$ is the Heaviside step function, $\sigma_0=\siginf(1-\eta_{se})$ is the DC conductivity, and subscript SE stands for stretched exponential. We want to obtain $\sigma_{se}$ from eq. \ref{eq: sigma_se_step_off}.
Taking the derivative with respect to time and multiplying by -1 yields:
\begin{equation}
    - \frac{\partial}{\partial t}\Big(\sigma_{se} \otimes (1-u_{step})\Big) = \sigma_{se}(t) \otimes \delta(t) = \sigma_{se}(t)
    \label{eq: sigma_se_impulse_general}
\end{equation}
where $\delta (t)$ is the Dirac-Delta function. Evaluating eq. (\ref{eq: sigma_se_impulse_general}) with eq. (\ref{eq: sigma_se_step_off}) results in
\begin{equation}
    \sigma_{se}(t)=\siginf \delta(t)-\siginf \eta_{se}t^{-1}\left(\frac{t}{\tau_{se}}\right)^{c_{se}}\rm{exp}\Big(-\left(\frac{t}{\tau_{se}}\right)^{c_{se}}\Big)u_{step}(t),
    \label{eq: sigma_se_impulse}
\end{equation}

A main reason why we used the SE conductivity function rather than the Cole-Cole function is its numerical advantage in the convolutional algorithm. With the SE conductivity, we do not need to convert  $\sigma(\omega)$ within each discretized voxel to $\sigma(t)$ because the SE conductivity has an explicit form in the time domain. The SE conductivity will not be beneficial when Maxwell's equations are solved in frequency-domain, and we believe that is the reason why the SE conductivity has not been used extensively, except for the latest simulation study from \cite{belliveau2018}.

\begin{figure}[htb]
  \centering
  \includegraphics[width=1.0\textwidth]{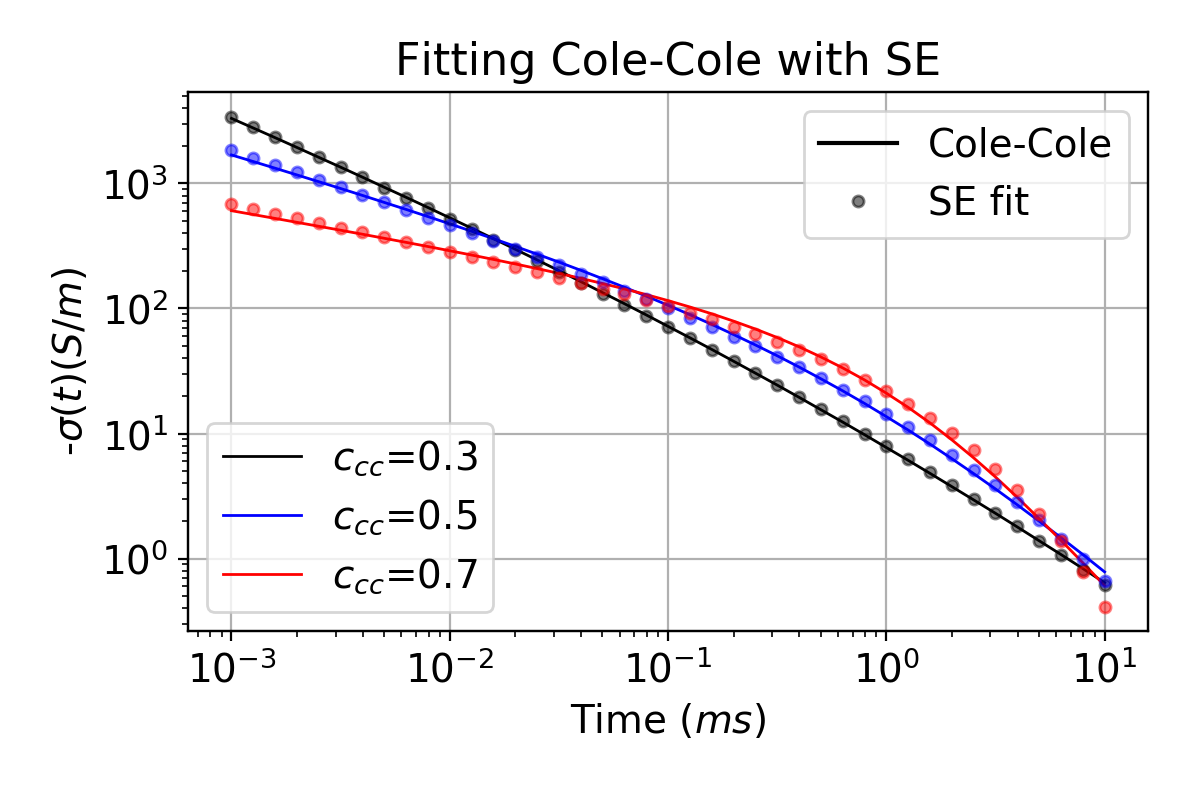}
  \caption{The stretched exponential (SE) fit of the Cole-Cole conductivity in the time domain. Solid lines are the impulse response using a Cole-Cole representation and the circles denote the SE response. Black, blue, and red colors correspondingly indicate when $c_{cc}$ is 0.3, 0.5, and 0.7.}
  \label{fig:1}
\end{figure}

Although the SE conductivity is not exactly the same as the Cole-Cole conductivity (eq. \ref{eq:colecole}), their time-features are very similar, and when $c_{cc}$=1 (Debye model), they are equivalent.
To illustrate cases when $c_{cc}$ is not equal to 1, we fit the Cole-Cole conductivity with the SE conductivity in time-domain; here, we update all three SE parameters: $\eta_{se}$, $\tau_{se}$, $c_{se}$ to fit Cole-Cole conductivity. Fig. \ref{fig:1} shows example Cole-Cole conductivity decays (t $>$ 0) with variable $c_{cc}$, and their fits with the SE conductivity.
For the range of times of interest (10$^{-3}$-10$^{1}$ms), the SE function effectively fits the Cole-Cole, as shown in Fig. \ref{fig:1}. They are essentially coincident. The estimated values of $\eta_{se}$ and $c_{se}$ are slightly smaller than their respective Cole-Cole counterparts; $\tau_{se}$ is coincident with $\tau_{cc}$ except when $c_{cc}$=0.2 (Table ~\ref{table: 1}). Therefore, when interpreting the SE parameters, readers can use their understanding of Cole-Cole parameters and treat the SE and CC parameters as being similar.
Note that we have used the impulse response of the Cole-Cole and SE functions when generating the fits. There is no loss of generality in doing this since the response due to an arbitrary waveform can be represented as a linear combination of impulse responses.

\begin{table}
 \centering
 \caption{Comparison of the Cole-Cole (CC) and the resultant Stretched Exponential (SE) parameters for variable $c_{cc}$. The chargeability, time constant, and the frequency component are correspondingly represented as $\eta$, $\tau$, $c$; subscripts CC and SE denote the Cole-Cole and SE parameters.}
 \begin{tabular}{@{}|c|c|c|c|c|c|c|}
    \hline
                &CC         &SE                &CC         &SE        &CC           &SE  \\
    \hline
    $\eta_{cc}$ &0.1        &0.09              &0.1        &0.09      &0.1          &0.09 \\
    $\tau_{cc}$ (ms) &1     &0.8               &1          &1         &1            &1 \\
    $c_{cc}$    &0.3        &0.2               &0.5        &0.4       &0.7          &0.6 \\
    \hline
 \end{tabular}
 \label{table: 1}
\end{table}
\clearpage

\section{NUMERICAL EXPERIMENTS}
\begin{figure}[htb]
  \centering
  \includegraphics[width=1.0\textwidth]{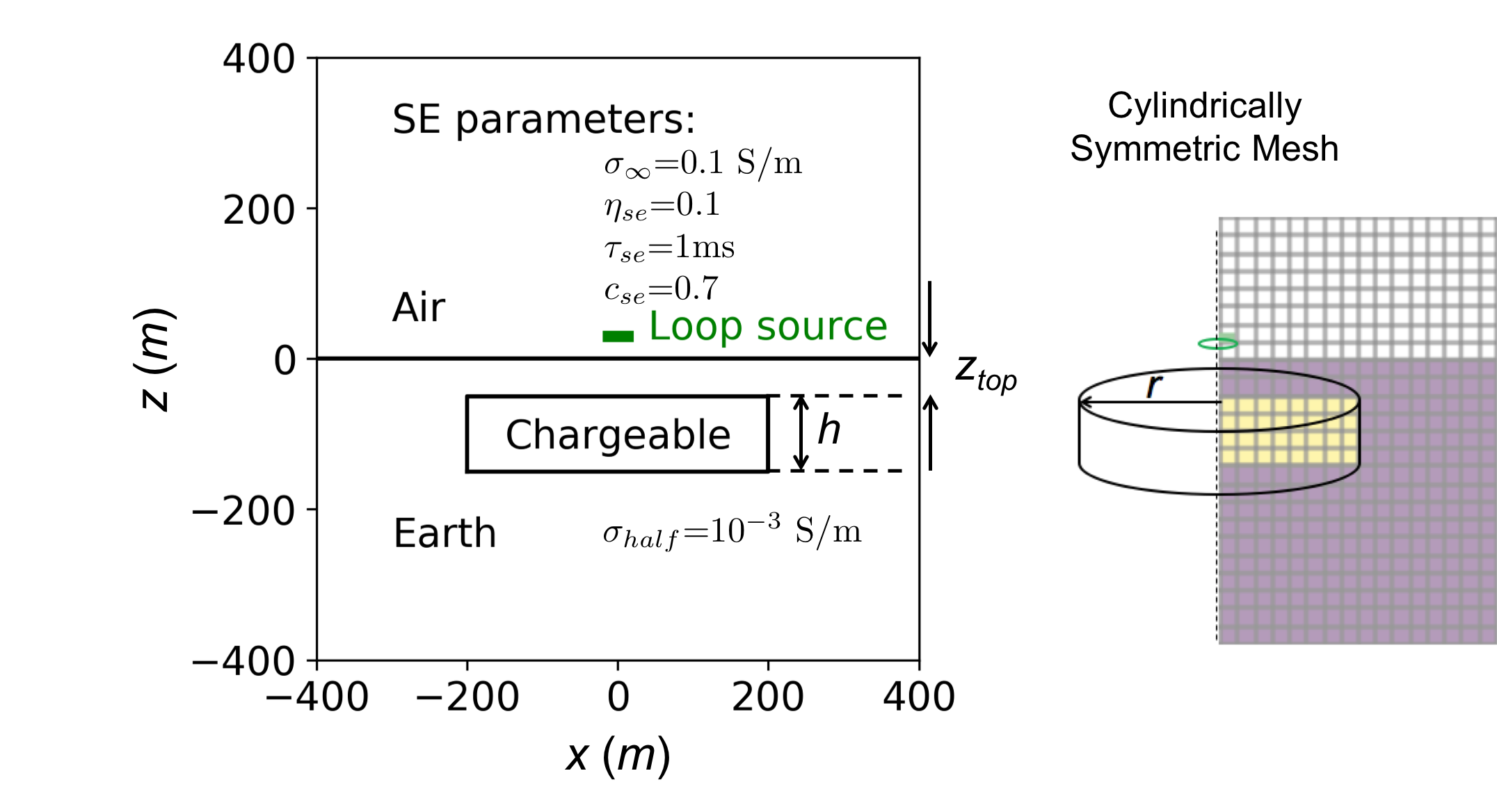}
  \caption{A chargeable cylinder embedded in a halfspace earth. The 13m-radius source loop is located 30 m above the surface. The depth to the top of the prism is denoted by $z_{top}$.  Right: the 2D cylindrically symmetric mesh is used for TEM simulations.}
  \label{fig:2}
\end{figure}

To answer the four questions posed previously, we carry out TEM simulations using the SimPEG-EMIP code. For the spatial discretization, we use the 2D cylindrically symmetric mesh because of the cylindrical symmetry in the time-domain AEM system, which uses a horizontal loop (See Fig. \ref{fig:2}). Rather than using a waveform for a specific AEM system, we use a step-off waveform as an input current. Again, there is no loss of generality since the response from an  arbitrary waveform can be generated by a linear combination of step-off responses \cite[]{fitterman1987}. A horizontal receiver loop measuring the voltage (equivalent to -$db_z/dt$) is coincident with the source loop. A chargeable cylinder is embedded in the resistive halfspace ($\sigma_{half}$=10$^{-3}$ S/m). The depth to the top ($z_{top}$), radius ($r$), and thickness ($h$) of the chargeable cylinder are correspondingly $z_{top}$=50 m, $r$=200 m, and $h$=100 m; the SE parameters of the cylinder are $\siginf$=0.1 S/m, $\eta_{se}$=0.1, $\tau_{se}$=1ms, $c_{se}$=0.7; the cylinder is 100 times more conductive than the halfspace and its effective conductance ($\sigma h$) is 10 S.

\begin{figure}[htb]
  \centering
  \includegraphics[width=1.0\textwidth]{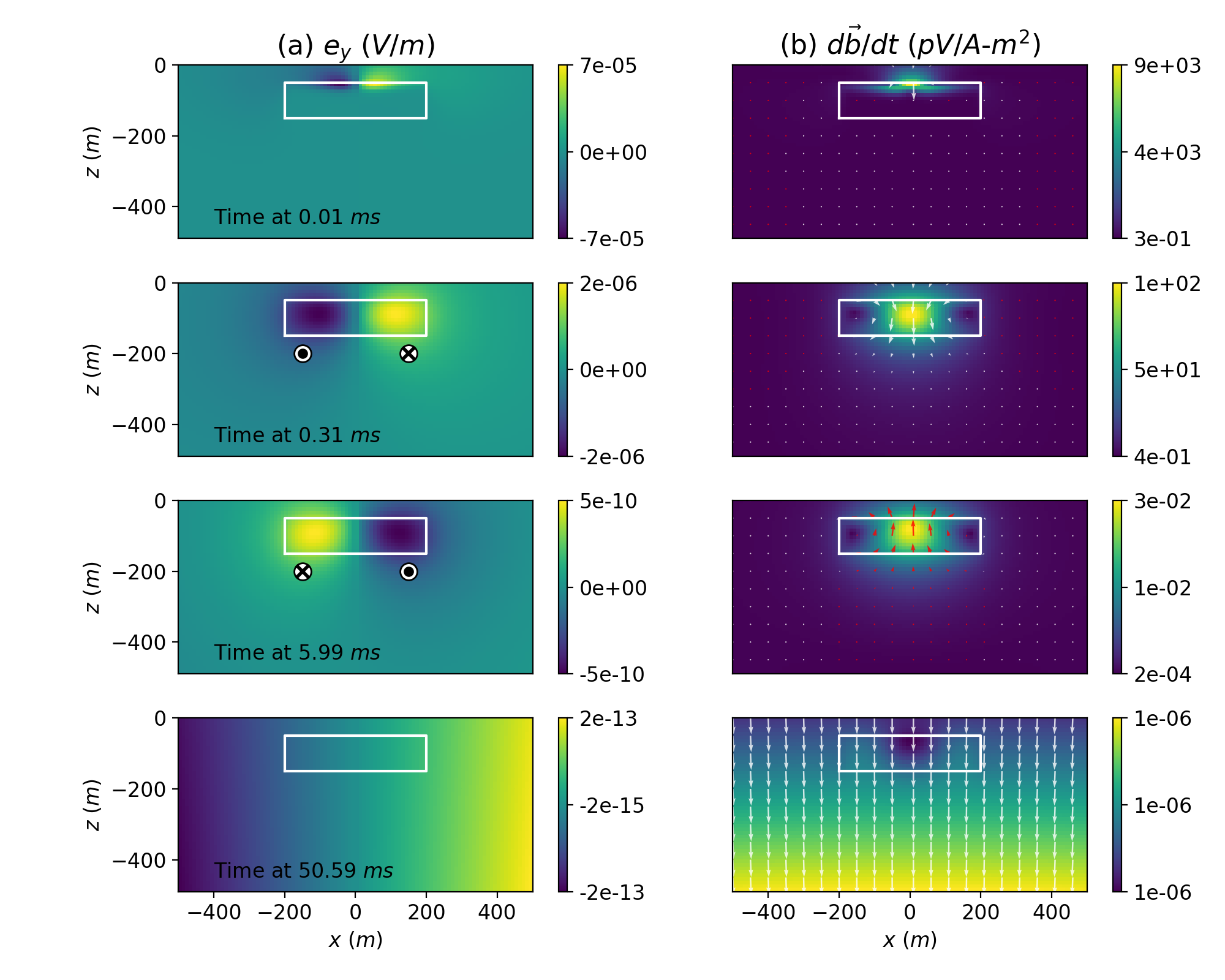}
  \caption{EM fields diffusing in the earth: (a) electric field in $y$-direction and (b) time derivative of the magnetic field ($d\b/dt$); the chargeable cylinder is outlined in white. At early times (0.01-0.3 ms), EM induction is dominant; this  results in inductive electric fields rotating counter-clockwise and $d\b/dt$ fields going upward. However, at 6ms the direction of the electric field is reversed (clockwise)  as a result of the chargeable cylinder; similarly $d\b/dt$ fields go upward (red arrows); This results in negative transients at the receiver loop (See Fig. \ref{fig:4}).}
  \label{fig:3}
\end{figure}

To understand how negative transients are caused by the presence of chargeable rocks, we first explore how the electric field diffuses into the earth after the input current is turned off.  Fig. \ref{fig:3}(a) shows the simulated electric field in y-direction (into the page) at four different time channels (0.01-50ms). At early times (0.01-0.3ms) electric fields, which rotate in the horizontal plane in a counter-clockwise direction, are induced in both the halfspace and in the conductive cylinder. As time passes, the electric field diffuses downwards and radially outwards; particularly large rotating electric fields are induced in the conductor. These inductive currents are responsible for ``charging up'' the earth material. At a later time (6 ms), the inductive currents have gone and only the decaying polarization currents remain. The resultant electric fields (and currents) have reversed direction; this is due to IP effects. At a later time (50ms), these IP effects have decayed away. With Faraday's law, electric fields are generated by time-varying magnetic field ($d\b/dt$), and the measured voltage is the same as the vertical component of -$d\b/dt$. Similarly, Fig. \ref{fig:3}(b) shows the vectoral distribution of $d\b/dt$ in time. At 0.3ms, the high amplitude of $d\b/dt$ is shown in the target, and the main direction of $d\b/dt$ (white arrow) is downward. However, at 6 ms the upward direction of $d\b/dt$ (red arrows) is generated by IP effects; this will result in negative transients at the receiver loop.
It is important to notice that electric fields generated either from EM or IP effects (Fig. ~\ref{fig:3}a)  do not cross a boundary. There is no charge build up on the boundary and there are no channeled currents (which is the mechanism by which IP signals for the grounded DC-IP surveys are generated). The current channeling  could happen if the cylindrical symmetry is broken (e.g. the source loop is located away from the center of the chargeable cylinder), but our analysis is focused on when cylindrical symmetry is preserved; the IP effects we show are solely due to the inductive polarized currents.

\begin{figure}[htb]
  \centering
  \includegraphics[width=1.0\textwidth]{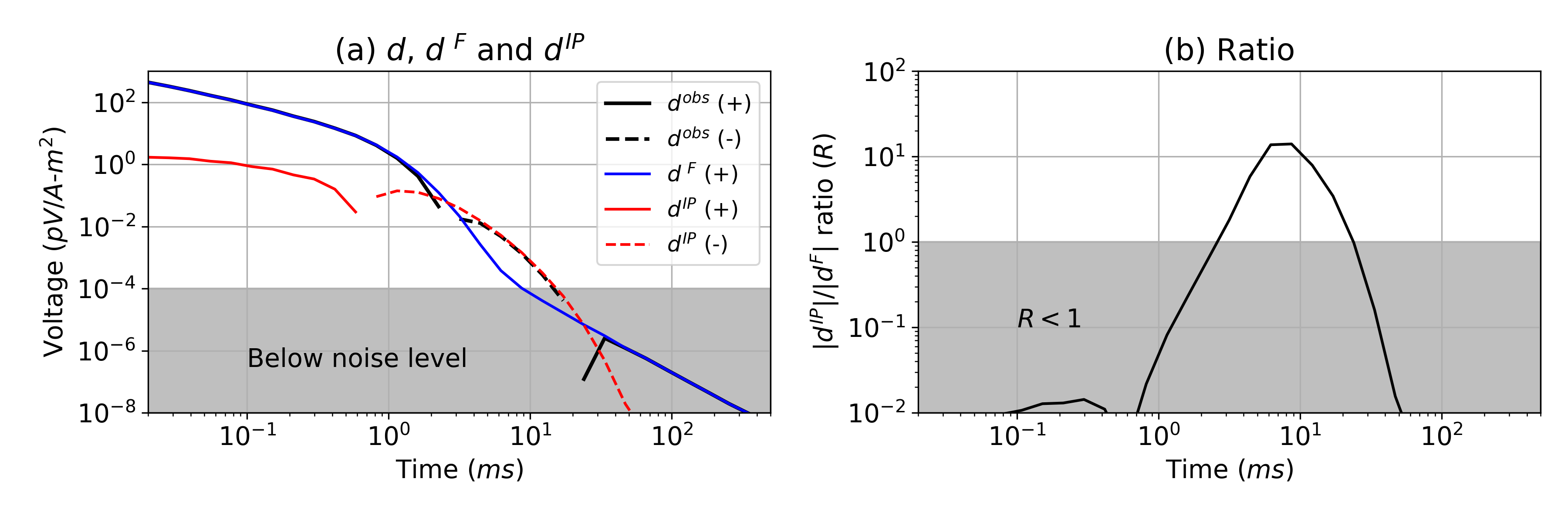}
  \caption{(a) Simulated TEM responses; $d^{obs}$ (black) are observations (EM+IP), $d^F$ (blue) are fundamental (EM) data, and $\dip$ (red) are IP data. Solid and dashed lines distinguish positive and negative values. Signals beneath the noise level (10$^{-4}$ pV/A-m$^2$) are shown within the grey zone. (b)  $R =\frac{|\dip|}{|d^F|}$ shows the relative strength of the IP signals compared to the fundamental induction effects. In the grey region  $R<1$ and the strength of the IP signal is smaller than the EM signal.}
  \label{fig:4}
\end{figure}

Based upon the physical understanding of IP effects due to a loop source, we examine  the data measured  at the receiver loop.
The black lines in Fig. \ref{fig:4}(a) show the measured time decays, $d^{obs}$, (on a log-log scale); negative values are shown after 2 ms (black dashed line). Another simulation is carried out without IP effects ($\eta_{se}$=0), and the computed data are shown with the blue line (no negatives); we call these the fundamental data, $d^F$; they include only EM induction effects. The IP data, $\dip$, are defined as
\begin{equation}
    \dip = d^{obs} - d^F.
\end{equation}
The system noise-level is set to 10$^{-4}$ pV/A-m$^2$ based upon a field data set measured at Mt. Milligan with a VTEM system (Figure 4.23 in \cite{Kang_2018}), which denoted as the grey shaded region in Fig. \ref{fig:4}(a). At the early times ($t<$1 ms), $d^{obs}$ and $d^{F}$ are almost coincident indicating that EM induction dominates the response. On the other hand, IP effects are dominant at later times ($t>$ 2ms). To show the relative strength of the IP effects, we define the ratio, $R$, between $|d^F|$ and $|\dip|$:
\begin{equation}
    R = \frac{|\dip|}{|d^F|}.
\end{equation}
We show $R$ in Fig. \ref{fig:4}(b); ratios smaller than 10$^{-2}$ are ignored. Between 1 ms and 40ms $R$ is greater than 0.1, indicating that there are considerable IP effects in the observations. When $R$=1, the observation is zero which corresponds to the time that the sign reversal occurs.

The EM induction processes within the background conductivity structure influence the electric field, which serves as a forcing function for IP effects. These IP effects translate into the reversed direction of the electric field in the target which results in negative transients that are observed after the EM induction effects have decayed. In the following sections, we carry out TEM-IP simulations  to systematically investigate the feasibility of the airborne IP technique. Variable model parameters are shown in Fig. \ref{fig:2}. Considering the typical  AEM system specifications, we limit our attention to the measured time range from 10$^{-2}$ms to 10$^{1}$ms and to voltages greater than the noise level (10$^{-4}$ pV/A-m$^2$). Further, for the metric pertaining to whether we can see IP signals or not, we use the existence of the negative datum being greater than the noise floor in the measured time range. Namely, we ignore subtle IP signals smaller than EM signals ($R<1$).

\clearpage
\subsection{Effects of background conductivity on IP signals}
As we demonstrated in Fig. \ref{fig:3}, the electric fields are the forcing functions which generate IP signals. Understanding the effects of the background conductivity on these electric fields is therefore crucial for understanding the resultant IP signals. To investigate this, we perform TEM-IP simulations for a range of values of $\siginf$ and $\sigma_{half}$. Other parameters for the simulation setup are the same as those in Fig. \ref{fig:1} ($r$=200m, $h$=100m, $z_{top}$=50m, $\eta_{se}$=0.1, $\tau_{se}$=1ms, $c_{se}$=0.7).

\begin{figure}[htb]
  \centering
  \includegraphics[width=1.0\textwidth]{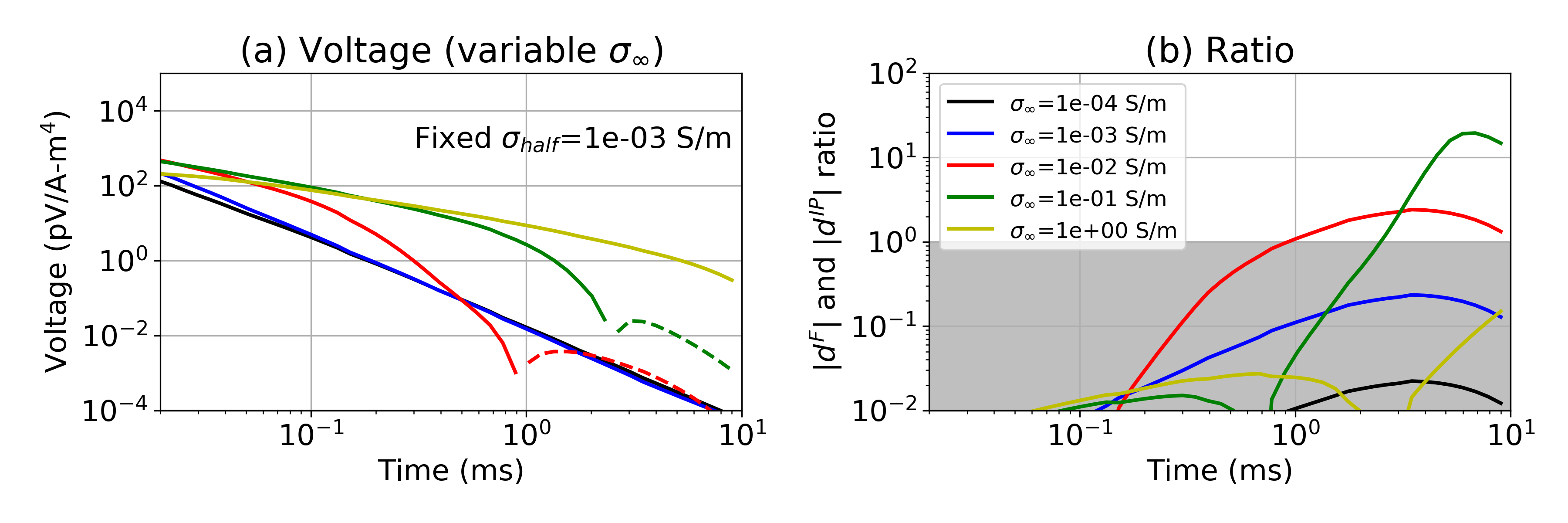}
  \caption{(a) Time decay curves with a variable conductivity of the chargeable cylinder.  Solid and dashed lines distinguish positive and negative values. The halfspace conductivity is fixed at 10$^{-3}$ S/m, whereas $\siginf$ varies (10$^{-4}$-0.1 S/m).  (b)  Plots of  $R=|\dip|/|d^F|$ . In the grey region $R<1$ and the strength of the IP signal is smaller than the EM signal. The legend for both plots is shown in (b). }
  \label{fig:5}
\end{figure}

The first experiment involves varying $\siginf$, which ranges from 10$^{-4}$ S/m to 1 S/m, while  $\sigma_{half}$ is fixed to 10$^{-3}$ S/m. Figs. \ref{fig:5}(a) and (b) show the simulated time decays and corresponding ratios, $R$. Negative transients are only visible when $\siginf$ is 0.01 S/m and  0.1 S/m (red and green curves). When $\siginf$ is too high (e.g. 1 S/m), EM effects dominate at all times and no negative transients are visible in the observed data. At the other end of the spectrum, the very resistive target shows the smallest $R$. These results show  that a moderately conductive target provides the best opportunity for observing negative transients in the data. When $\siginf$=1 S/m, there are no negatives in the time decay curve.

\begin{figure}[htb]
  \centering
  \includegraphics[width=1.0\textwidth]{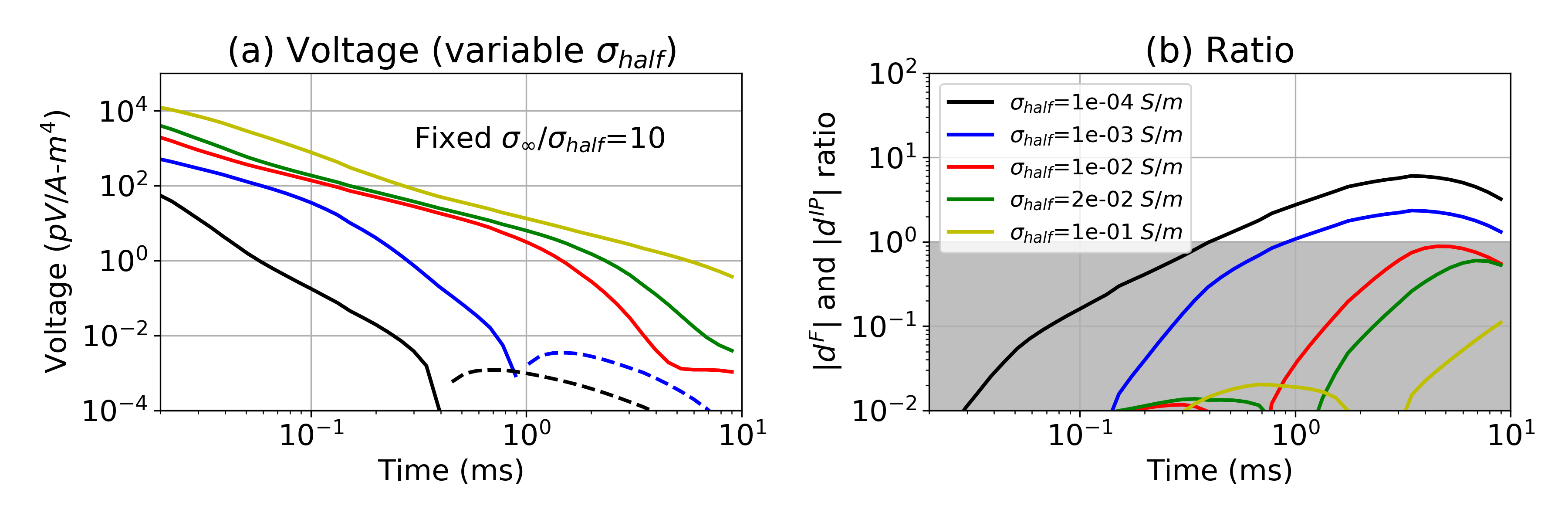}
  \caption{(a) Time decay curves with a variable conductivity of the halfspace, $\sigma_{half}$.  Solid and dashed lines distinguish positive and negative values. The ratio between the halfspace conductivity and the target conductivity ( $\siginf$/$\sigma_{half}$), is fixed to 10, whereas  $\sigma_{half}$ varies (10$^{-4}$-0.1 S/m). (b)  Plots of the $|\dip|/|d^F|$ ratio, $R$. In the grey region $R<1$ and the strength of the IP signal is smaller than the EM signal. The legend for both plots is shown in (b).}
  \label{fig:6}
\end{figure}

To explore the effect of the halfspace conductivity, $\sigma_{half}$, we fix the ratio $\siginf$/ $\sigma_{half}$ to be 10, and change $\sigma_{half}$ from 0.1 S/m to 10$^{-4}$ S/m. In Fig. \ref{fig:6}, negatives are present when $\sigma_{half}$  is 10$^{-4}$ S/m and 10$^{-3}$ S/m, but not for the other cases. This shows that when the conductivity of the non-chargeable halfspace is too high ($>$0.1 S/m), measuring IP signals will be challenging even though chargeable materials exist. Therefore, a moderately conductive target ($\sim$0.01-0.1 S/m) in a resistive host ($\sim$10$^{-4}$-10$^{-3}$ S/m) provides the best circumstances for observing strong IP signals.

\clearpage

\subsection{To what depth can we expect to detect chargeable material with airborne IP?}
\begin{figure}[htb]
  \centering
  \includegraphics[width=1.0\textwidth]{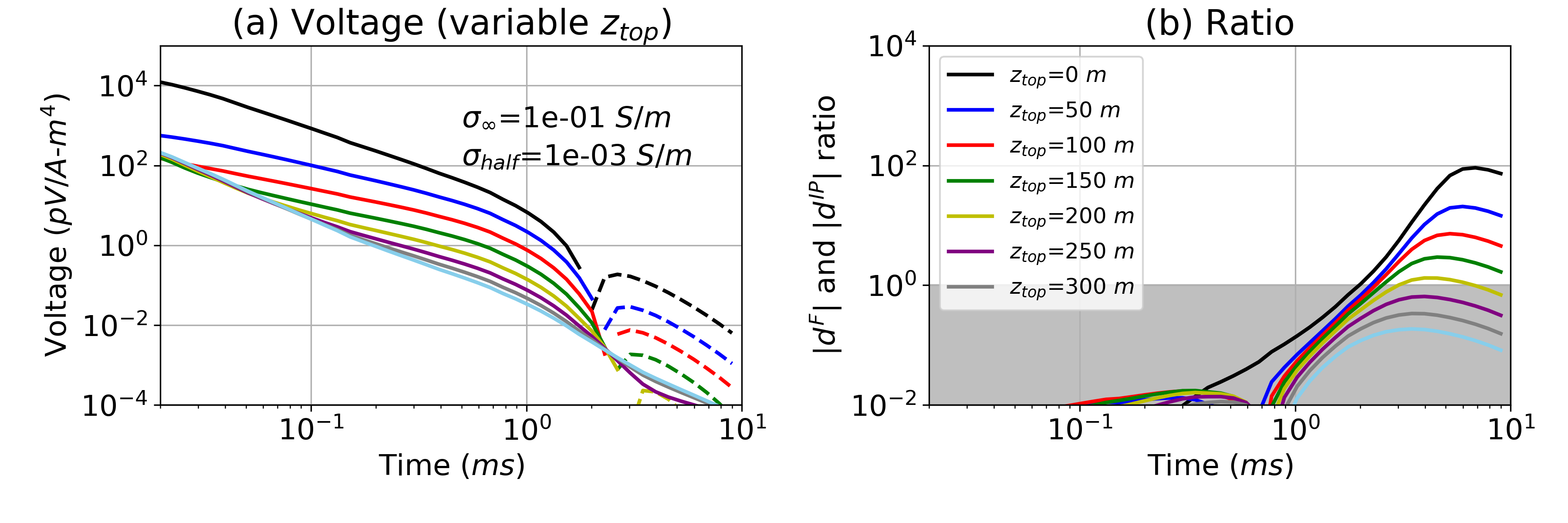}
  \caption{(a) Time decay curves with a variable target depth ($z_{top}$) ranging from 0-350 m. The halfspace conductivity is 10$^{-3}$ S/m. (b)  Plots of  $R=|\dip|/|d^F|$ . In the grey region $R<1$ and the strength of the IP signal is smaller than the EM signal. The legend for both plots is shown in (b).}
  \label{fig:7}
\end{figure}

\begin{figure}[htb]
  \centering
  \includegraphics[width=1.0\textwidth]{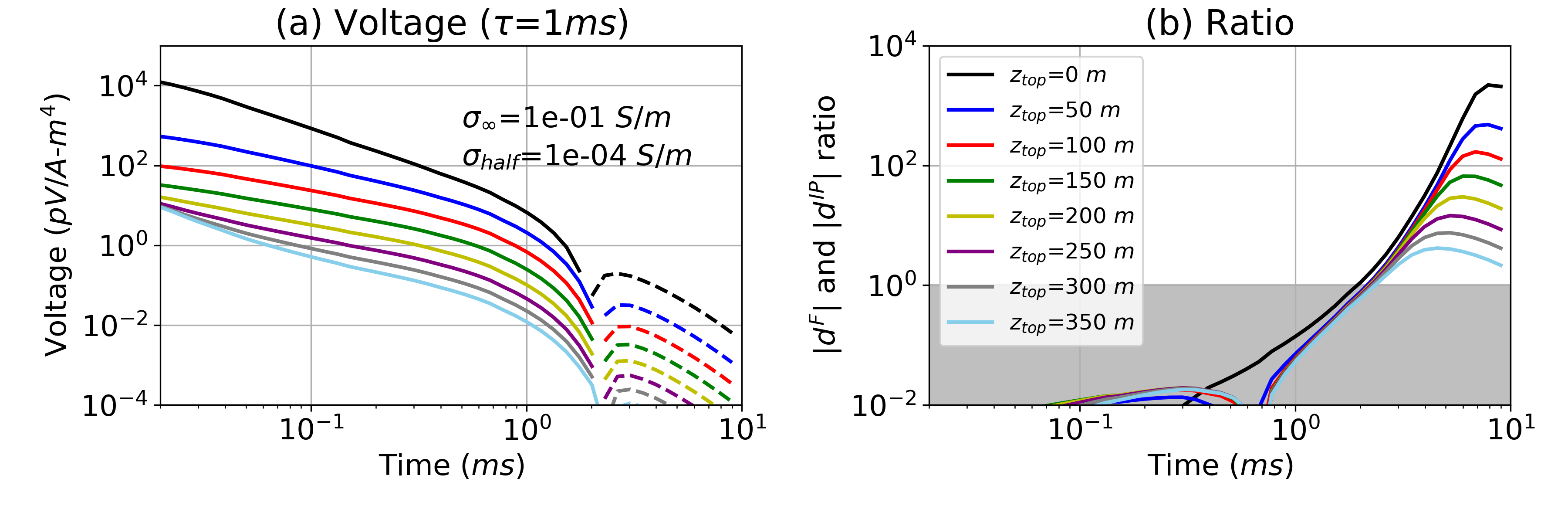}
  \caption{(a) Time decay curves with a variable target depth ($z_{top}$) ranging from 0-350 m.  The halfspace conductivity is decreased to 10$^{-4}$ S/m compared to Fig. \ref{fig:7}. (b)  Plots of  $R=|\dip|/|d^F|$ . In the grey region $R<1$ and the strength of the IP signal is smaller than the EM signal. The legend for both plots is shown in (b).}
  \label{fig:8}
\end{figure}

Often, the maximum depth that airborne IP can see chargeable targets is considered to be fairly low ($z<\sim$100 m from \cite{macnae2016}). However, the possibility exists to see deeper when the host is resistive and the chargeable target is moderately conductive. Here we explore detectability of a chargeable target by altering the depth of the target ($z_{top}$) from 0m to 350 m. Fig. \ref{fig:7} shows the time decays with variable $z_{top}$ when  $\siginf$ and $\sigma_{half}$ are 0.1 S/m and 10$^{-3}$ S/m respectively. Negatives are present when $z_{top}$ $\leq$ 200m. By decreasing $\sigma_{half}$, this depth can be increased to 300 m as shown in Fig. \ref{fig:8}. Hence, it is possible to detect a deeper chargeable target using the airborne IP technique when the target is moderately conductive and the host rock is resistive (10$^{-4}$ S/m). For instance, at the Tli Kwi Cho kimberlite deposit, negatives were measured near a kimberlite pipe. This moderately conductive pipe was embedded in a resistive host rock (10$^{-4}$ S/m); it was located  $\sim$70 m below the surface and its radius and thickness were approximately 150 m and 200 m, respectively \cite[]{kang2017}. This geometry is  similar to that of our chargeable cylinder shown in Fig. ~\ref{fig:2}.

\begin{figure}[htb]
  \centering
  \includegraphics[width=1.0\textwidth]{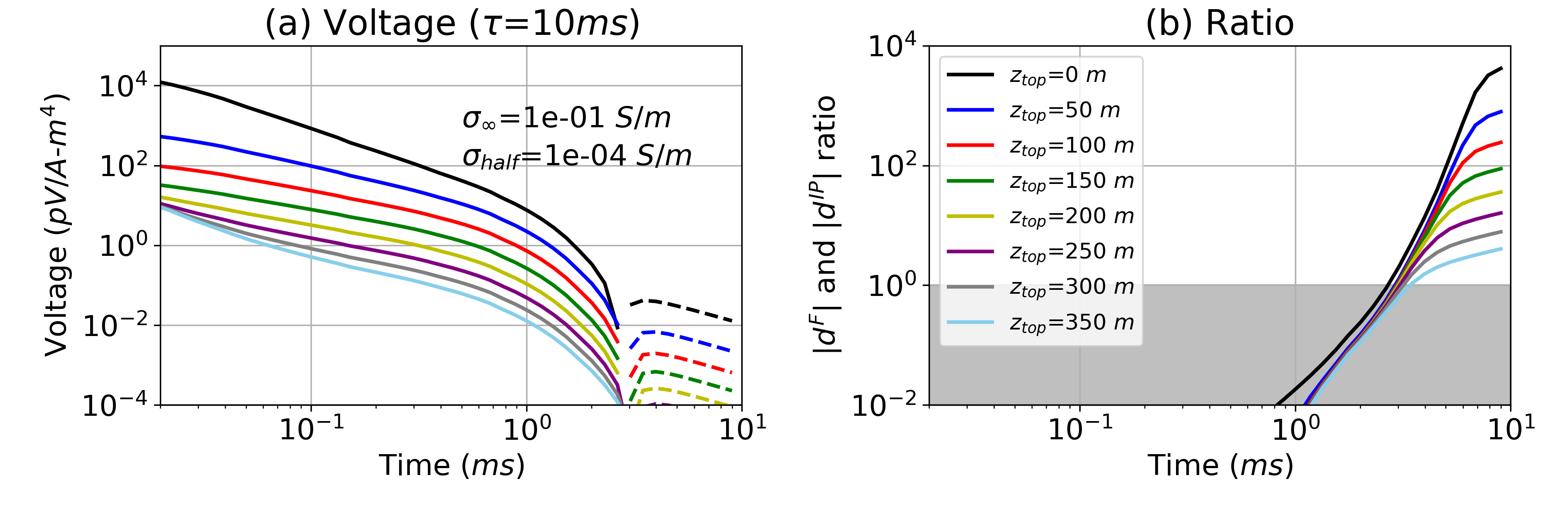}
  \caption{(a) Time decay curves with a variable target depth ($z_{top}$) ranging from 0-350 m.  The time constant ($\tau$) is increased to 10$^{-2}$s. (b) $|\dip|/|d^F|$ ratio, $R$. Grey region indicates $R<1$ meaning the strength of the IP signal is smaller than EM signal.}
  \label{fig:8-1}
\end{figure}

\begin{figure}[htb]
  \centering
  \includegraphics[width=1.0\textwidth]{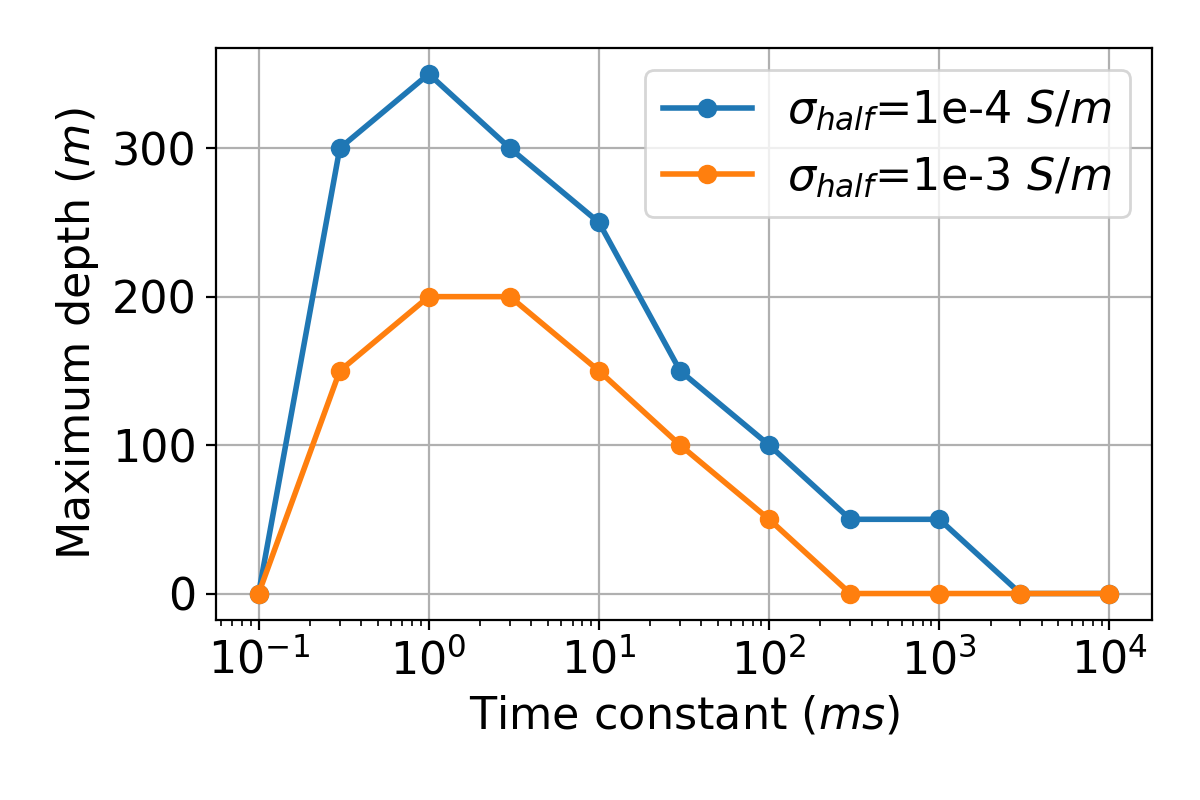}
  \caption{Maximum depth of the chargeable cylinder at which we can observe negative transients as a function of the time constant (ms). Light blue and orange lines indicate two conductivity models having different halfspace conductivity ($\sigma_{half}$): 10$^{-4}$ S/m and 10$^{-3}$ S/m, respectively.}
  \label{fig:8-2}
\end{figure}

The maximum depth that we can see negatives will depend upon IP parameters. For instance, greater chargeability will increase the strength of the IP signals and therefore the maximum depth can be increased with increased chargeability \cite[]{macnae2016}. The effects of the time constant are more complicated to unravel. We explore this by changing the  time constant of the target from 1ms to 10ms and altering the depth of burial.  We obtain Fig. ~\ref{fig:8-1} and observe that the maximum depth is decreased from 350m to 250m. Performing a similar analyses for the time constant ranging from  0.1ms to 10s, we obtain the maximum depth as a function of the time constant as shown in Fig. ~\ref{fig:8-2}. The maximum depth starts from zero when $\tau_{se}$=0.1ms, increases until $\tau_{se}$=1ms, and  then decreases as $\tau_{se}$ increases. We simulated two cases in which  $\sigma_{half}$ was 10$^{-3}$ S/m and 10$^{-4}$ S/m but the conductivity of the target ($\siginf$) was fixed to 10$^{-1}$ S/m. Greater maximum depth is shown when $\sigma_{half}$=10$^{-4}$ S/m.
Hence there is an optimal time constant ($\sim$1ms) that can generate the greatest IP signals. This can be understood from the following. Considering the measured time range of the data: 10$^{-2}$-10ms, there simply not enough time to charge up material that has time constant greater than 3s. Further, when the IP decay is too fast (small time constant) compared to EM decay, it is hard to be the signal in the observation.

\clearpage

\subsection{The effects of target size}
\begin{figure}[htb]
  \centering
  \includegraphics[width=1.0\textwidth]{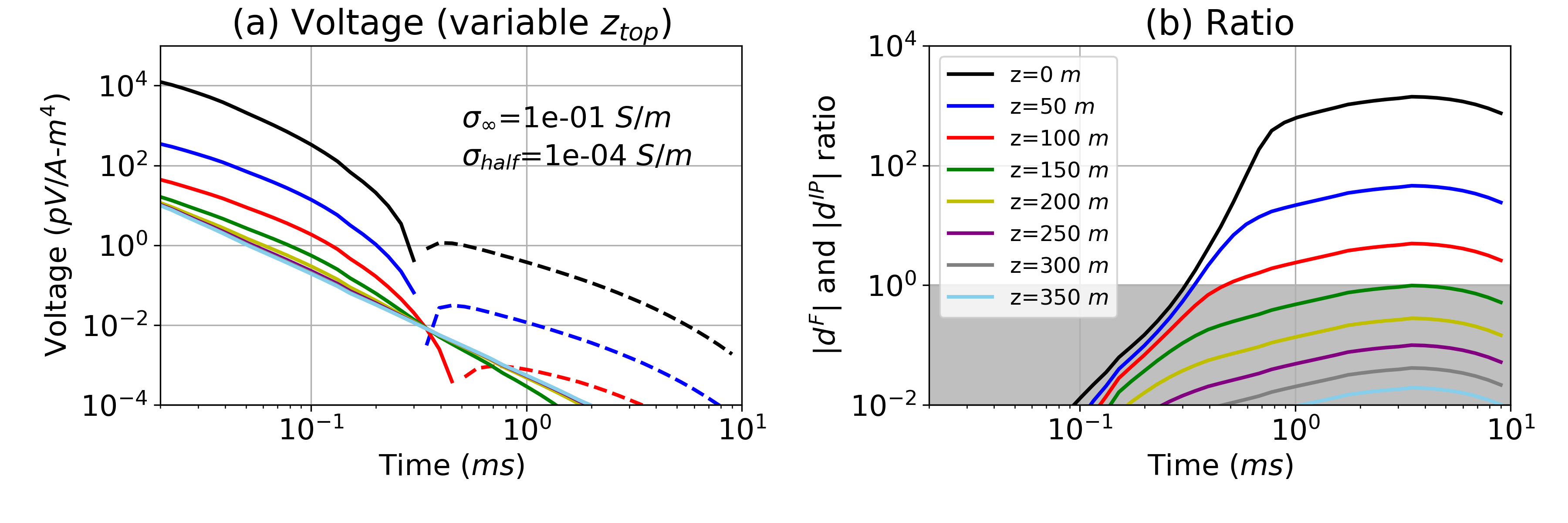}
  \caption{(a) Time decay curves with a variable target depth ($z_{top}$). The radius ($r$) of the chargeable cylinder is decreased from 200 m to 100 m compared to Fig.  8. (b) $|\dip|/|d^F|$ ratio, $R$. Grey region indicates $R<1$ meaning the strength of the IP signal is smaller than EM signal.}
  \label{fig:9}
\end{figure}

\begin{figure}[htb]
  \centering
  \includegraphics[width=1.0\textwidth]{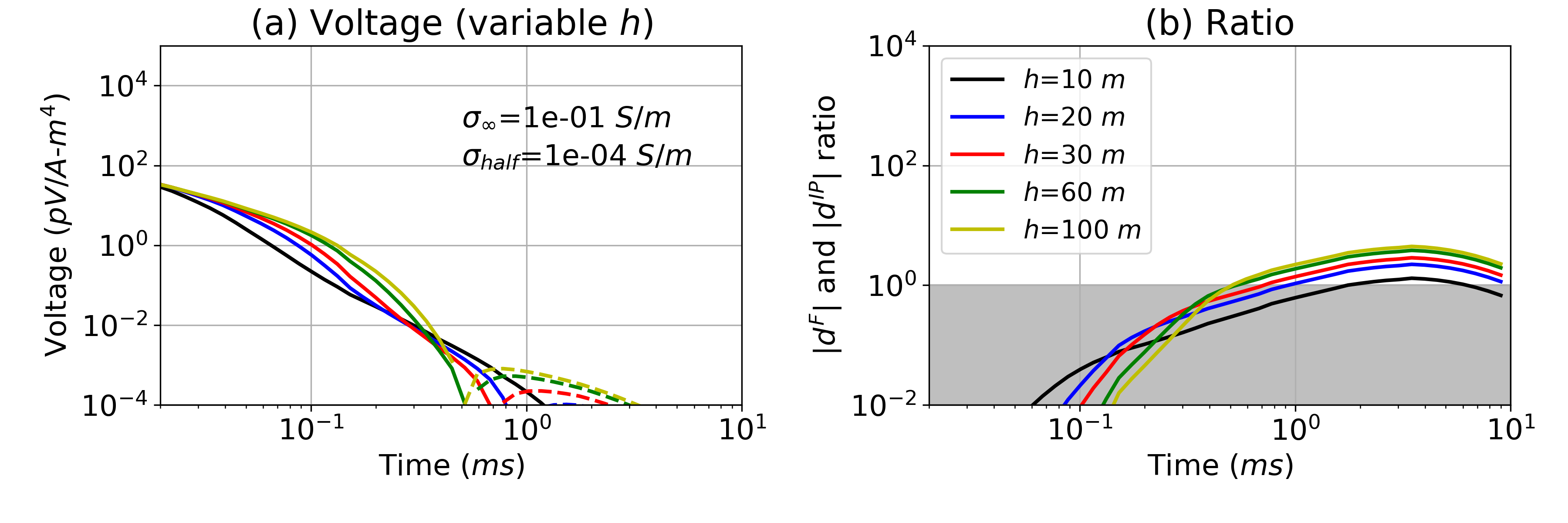}
  \caption{(a) Time decay curves with a variable target thickness (h). The radius ($r$) and depth to the top ($z_{top}$) of the chargeable cylinder are fixed to 50 m and 150 m, respectively, whereas $h$ varies (10-100 m) as shown in the legend in (b) which displays the $|\dip|/|d^F|$ ratio. Grey region indicates $R<1$ meaning the strength of the IP signal is smaller than EM signal.}
  \label{fig:10}
\end{figure}

The examples shown so far have illustrated general principles concerned with the ability to detect IP bodies at depth. We have dealt with a specific geometry and have worked with a fairly large target. The strength of the IP signal depends upon the size and geometry of the target. To begin an exploration of the impact of target size on detectability, we show the effects of making the body smaller. We first reduce the radius ($r$) from 200 m to 50 m. As a result, the maximum depth has decreased from 300m (Fig. ~\ref{fig:8}) to 100 m as shown in Fig. \ref{fig:9}. We can also reduce the thickness ($h$), as shown in Fig. \ref{fig:10}; here $z_{top}$=150m and $r$=100m. When $h$=10m, we no longer observe the negatives. Depending on the geological setting, more complicated situations may occur and the potential for seeing an IP signal in the airborne data will require 3D modelling appropriate to the geology. For instance, layering of the subsurface can also make significant impact to the maximum depth in practice, and this was not taken account in our analyses.

\clearpage

\subsection{Which are the characteristics of  detectable chargeable materials in AEM data?}
\begin{figure}[htb]
  \centering
  \includegraphics[width=1.0\textwidth]{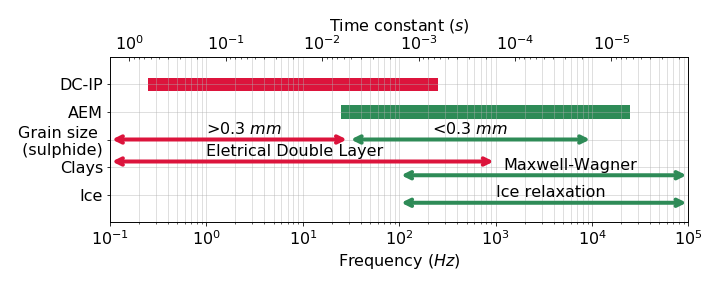}
  \caption{Frequency spectrum of EM systems (DC-IP and AEM) and various IP sources (sulphide, clays, and ice). The time constant is denoted at the top $x$-axis. This figure is based upon previous research \cite[]{pelton1978,Revil2013a,grimm2015,macnae2016b}}.
  \label{fig:11}
\end{figure}

The different polarization characteristics of earth materials can be translated into different Cole-Cole or SE parameters. In particular, their frequency spectrum, or the time period over which the polarization process is occurring, differs; this time period is closely related to the time constant, $\tau$. For instance, fine-grained sulphides will show a higher frequency spectrum and smaller time constant than coarse-grained sulphides. Ground and airborne surveys such as DC-IP and AEM often use different base frequencies: e.g. 0.125 Hz and 25 Hz. As AEM surveys have a higher base frequency, they are more sensitive to chargeable materials which are characterized by high frequency or smaller time constants. Three main chargeable targets of interest in airborne IP are (a) fine-grained sulphides \cite[]{pelton1978,Revil2017}, (b) clays \cite[]{macnae2016,leroy2009}, and (c) ice \cite[]{grimm2015,kang2017}. In Fig. \ref{fig:11}, we have plotted the frequency spectrum of each of these materials along with the frequency spectrum of DC-IP and AEM surveys.

\begin{figure}[htb]
  \centering
  \includegraphics[width=1.0\textwidth]{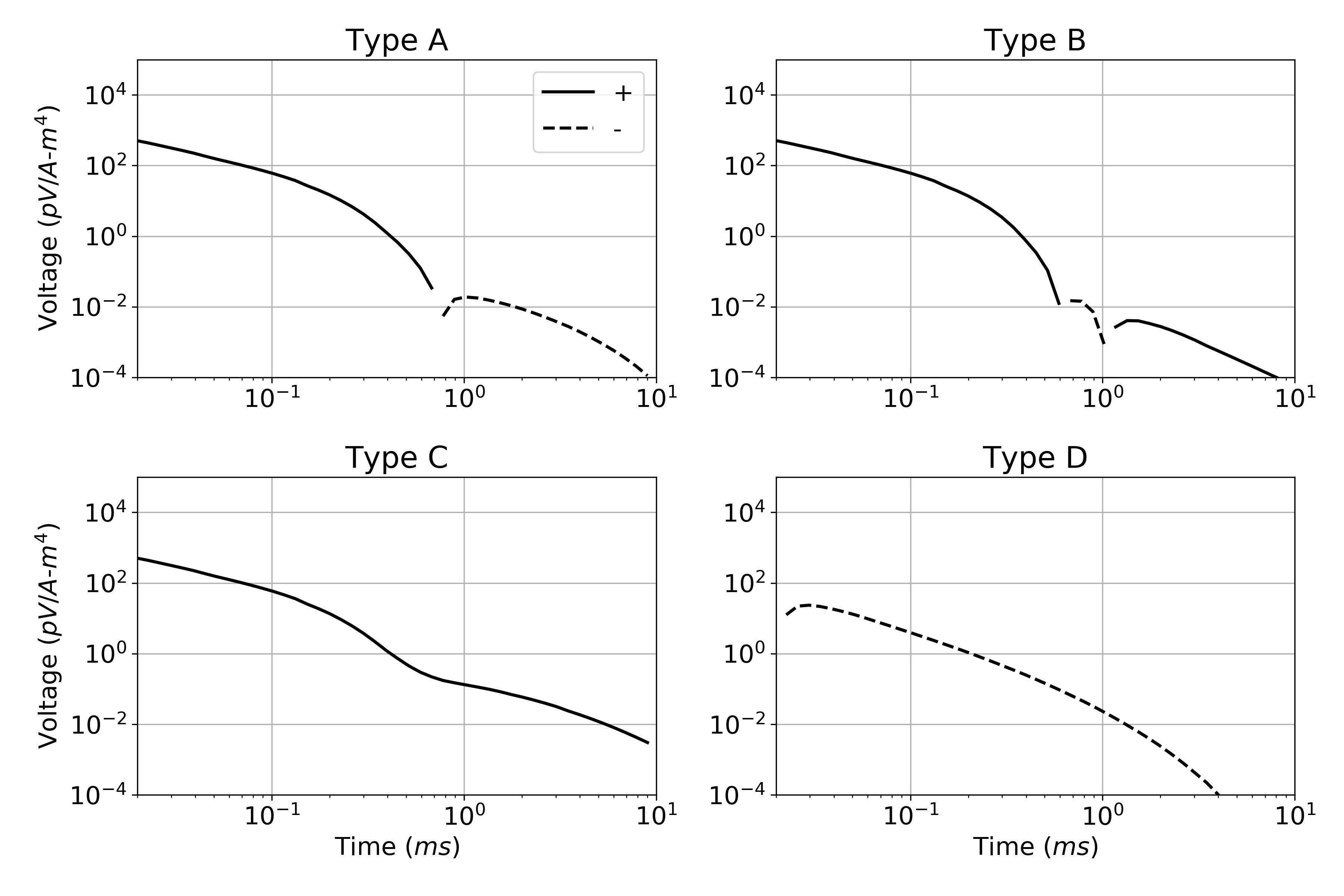}
  \caption{Four different types of time decays (A-D) from different sources of IP. Parameters used to compute time decays are summarized in Table \ref{table: 2}.}
  \label{fig:12}
\end{figure}

To examine how each of these materials impacts the observed time-decays, we have defined four models in Table \ref{table: 2} for which we will simulate AEM data. The decays, plotted in Fig. \ref{fig:12}, show the characteristic time behavior associated each of the chargeable materials:

\begin{itemize}
    \item Type A: Typical time decay showing positive early time data and negative late time data; this can be generated by fine-grained sulphides and clays.
    \item Type B: Double sign-reversal; when sulphides or clays have very fine grain size the resulting time constant is smaller.
    \item Type C: No negatives, but a positive `bump' at late time; this is when there is a deep conductor below a chargeable target, which can generate strong positive EM signals at late time.
    \item Type D: No positives; this can be generated by an extremely chargeable target such as ice ($\simeq$0.9) located very near surface, or not measuring early enough time channels.
\end{itemize}

\begin{table}
 \centering
 \caption{Parameters of the chargeable cylinder for the type curves (A-D) shown in Fig. 9. For Type C a conductive layer (0.1 S/m) is added 300 m below the surface; its thickness is 100 m.}
 \vspace{0.5cm}
 \begin{tabular}{@{}|c|c|c|c|c|}
    \hline
    Division    &Type A &Type B &Type C & Type D\\
    \hline
    Lithology    &\makecell{Clay \\ Sulphide} & \makecell{Clay (finer) \\ Sulphide (finer)} & \makecell{Type A \\ with a deep conductor} & Ice \\
    $z_{top}$ (m)                 &50 &50 &50 & 0\\
    $\sigma_{half}$ (S/m)   &10$^{-3}$ &10$^{-4}$ &10$^{-3}$ & 10$^{-4}$\\
    $\siginf$ (S/m)         &2$\times$10$^{-2}$ &2$\times$10$^{-2}$ &2$\times$10$^{-2}$ & 10$^{-3}$\\
    $\eta_{se}$             &0.1 &0.1 &0.1 & 0.9\\
    $\tau_{se} (ms)$        &1 &0.1 &1 & 0.08\\
    $c_{se}$                &0.7 &0.7 &0.7 & 0.5 \\
    \hline
 \end{tabular}
 \label{table: 2}
\end{table}
\clearpage

\section{Conclusions}
We have developed a convolutional TEM simulation code that directly solves Maxwell's equations in time with the SE (stretched exponential) conductivity function. The SE conductivity is a good representation of the Cole-Cole conductivity for the typical time range used in AEM. With our simulations, we showed that:
\begin{itemize}
    \item Negative transients in AEM data can be caused by the reversed direction of the electric field in a chargeable target and are visible at late times when EM induction is small.
    \item Moderately conductive targets (0.01-0.1 S/m) in a resistive host (10$^{-4}$-10$^{-3}$ S/m) show the best potential for generating strong IP signals (negatives) in AEM data.
    \item The depth at which we can detect a target with airborne IP depends upon the background conductivity, but for an ideal situation (a conductive, chargeable target in a resistive host), the target can be detected up to 300 m depth provided its time constant is close to 1ms. This maximum depth will naturally be affected by the layering, but this was not taken account in our analyses.
    \item The strength of the IP signals and the depth of detectability of a target is dependent upon the size of the target and its geometry. In general 3D simulations, cast within the relevant geologic context, are required.
    \item The three main sources of chargeable material detectable in AEM (fine grain sulphides, clay and ice) can give rise to four different characteristic decay curves.
\end{itemize}

The overriding question of practical concern is whether, or under what circumstances, you can see an IP target at depth in airborne EM data. The situation is complex and cannot be answered by a simple, fixed depth of investigation rule. Forward simulation which emulates the potential geology and its associated physical properties, is required. To advance this capability, we have developed the SimPEG-EMIP codes as a part of the open-source software project, SimPEG (\url{https://www.simpeg.xyz}). The main workhorse for this paper has been the 2D cylindrical mesh, and the figures in this text can be reproduced with provided Jupyter Notebooks (\url{https://github.com/simpeg-research/kang-2018-AEM}). Moreover, the source code is downloadable and users can explore the use of different parameters. We hope these efforts contribute to the wider challenge of extracting IP information from airborne time-domain EM data.

\append{Discretization}
\label{app: Discretization}
In this section, we discuss important elements about discretizing Maxwell’s equations in the time-domain with the convolution term shown in eq. (\ref{eq:ohmslaw_time}), to simulate IP effects in time-domain EM data. Appendix A.1 illustrates how convolutionary time-domain Maxwell's equations can be discretized.  Appendix A.2 describes how the singularity of SE conductivity function at $t=0$ is handled. Most of key challenges about this discretization are tackled in \cite{marchant2015} (see page 21), and we have extended his work, applied for Cole-Cole conductivity, to SE conductivity.

\subsection{Maxwell's equations}
\label{app: Maxwell's equations}
\begin{figure}[htb]
  \centering
  \includegraphics[width=1.0\textwidth]{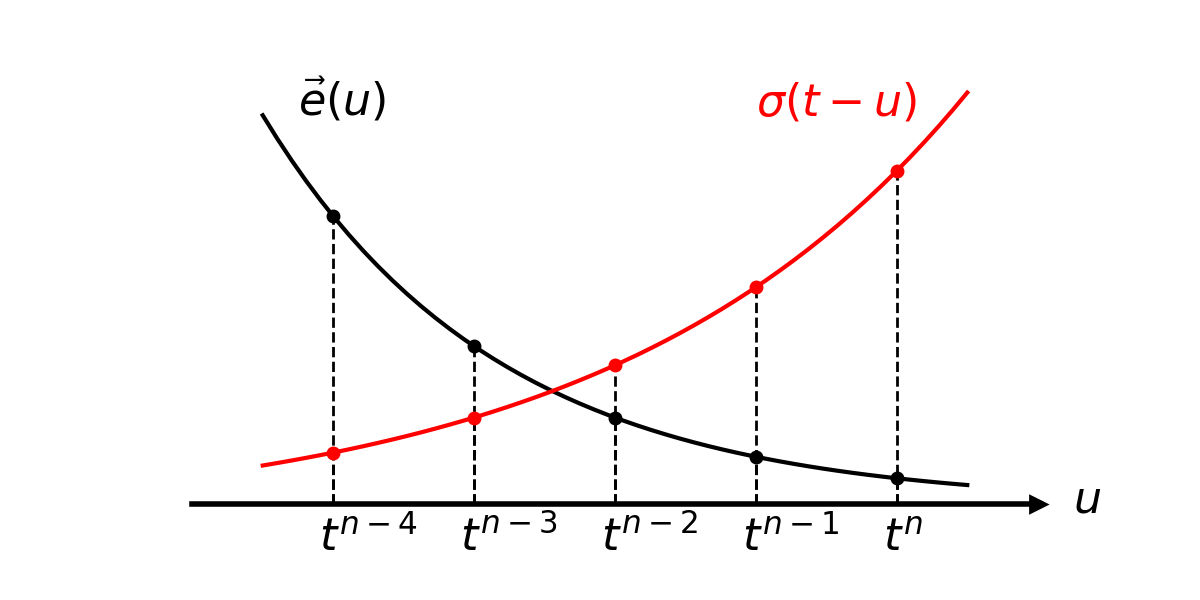}
  \caption{Conceptual diagram to describe discrete convolution process in eq. (\ref{eq:convolution_trapezoidal})}
  \label{fig:convolution_concept}
\end{figure}

The stretched exponential (SE) conductivity provided in eq. (\ref{eq: sigma_se_impulse}) in the time-domain can be rewritten as
\begin{equation}
  \sigma_{se}(t) = \siginf \delta (t) + \dsig (t)
  \label{eq:sigma_time}
\end{equation}
where  $\delta(t)$ is a Dirac-Delta function and $\dsig(t)$ is
\begin{equation}
  \dsig (t) = -\siginf \eta_{se}t^{-1}(\frac{t}{\tau_{se}})^{c_{se}}exp\Big(-(\frac{t}{\tau_{se}})^{c_{se}}\Big)
\end{equation}
Considering a time-dependent conductivity, Ohm's Law can be written as
\begin{equation}
  \j = \sigma_{se}(t) \otimes \e =\int_0^t \sigma(t-u) \e (u) du
  \label{eq:ohmslaw}
\end{equation}
and substituting eq. (\ref{eq:sigma_time}) yields
\begin{equation}
  \j = \siginf \e + \int_0^t \dsig(t-u) \e (u) du
  \label{eq:ohmslaw_two}
\end{equation}
Using the Backward Euler method, we discretize Maxwell's equations in eqs. (\ref{eq:faraday}) and (\ref{eq:ampere}) in time:
\begin{equation}
  \curl \e^{\ (n)} = -\frac{\b^{(n)}-\b^{(n-1)}}{\triangle t^{(n)}}
  \label{eq:faraday_time}
\end{equation}
\begin{equation}
  \curl \mu^{-1} \b^{(n)} - \j^{(n)} = \j_s^{(n)} \\
  \label{eq:ampere_time}
\end{equation}
where $\triangle t^{(n)} = t^{(n)}- t^{(n-1)}$.
To discretize the integral in eq. (\ref{eq:ohmslaw_two}), we use the trapezoidal rule:
\begin{equation}
  \int_{t^{(k-1)}}^{t^{(k)}} \dsig(t-u) \e (u) du
  = \frac{\triangle t^{(k)}}{2} \Big(\dsig (t^{(n)} - t^{(k-1)}) \e^{\ (k-1)} + \dsig (t^{(n)} - t^{(k)}) \e^{\ (k)} \Big)
  \label{eq:convolution_trapezoidal}
\end{equation}
Fig. \ref{fig:convolution_concept} shows a conceptual diagram for this discrete convolution procedure.
Hence eq. (\ref{eq:ohmslaw_two}) can be discretized as
\begin{equation}
  \j^{(n)} = \siginf \e^{\ (n)} +
  \sum_{k=1}^{n} \frac{\triangle t^{(k)}}{2} \Big(\dsig (t^{(n)} - t^{(k-1)}) \e^{\ (k-1)} + \dsig (t^{(n)} - t^{(k)}) \e^{\ (k)} \Big)
  \label{eq:ohmslaw_time_original}
\end{equation}
This can be rewritten as
\begin{equation}
  \j^{(n)} = \Big(\siginf + \gamma (\triangle t^{(n)})\Big)\e^{\ (n)} + \j_{pol}^{(n-1)}
  \label{eq:ohmslaw_time}
\end{equation}
where the polarization current, $\j_{pol}^{(n-1)}$ is
\begin{align}
  \j_{pol}^{(n-1)} = \sum_{k=1}^{n-1} \frac{\triangle t^{(k)}}{2} \Big(\dsig (t^{(n)} - t^{(k-1)}) \e^{\ (k-1)} + \dsig (t^{(n)} - t^{(k)}) \e^{\ (k)} \Big) \nonumber \\
  +  \kappa(\triangle t^{(n)}) \e^{\ (n-1)}
\end{align}
For the simplest case when ($c_{se}=1$), then $\dsig(t=0)$  is well defined and  $\gamma (\triangle t^{(n)})$ and $\kappa (\triangle t^{(n)})$ are respectively:
\begin{equation}
  \gamma(\triangle t^{(n)}) = \frac{\triangle t^{(n)}}{2}\dsig (0),
\end{equation}
\begin{equation}
  \kappa(\triangle t^{(n)}) = \frac{\triangle t^{(n)}}{2} \dsig (\triangle t^{(n)})
\end{equation}
However, when $c_{se}\neq1$, $\dsig(t=0)$ is singular and hence it requires special numerical treatment; this is described in Appendix A.2.

For the discretization, we use a staggered mimetic finite volume approach \cite[]{hyman2002}. Here, boldface with uppercase and lowercase indicate matrices and column vectors, respectively. Further details about the discretization can be found in \cite{haber_book} (see page 31).
Discretizing eqs. (\ref{eq:faraday_time}), (\ref{eq:ampere_time}), and (\ref{eq:ohmslaw_time}) yields
\begin{equation}
  \dcurl \de^{\ (n)} = -\frac{\db^{(n)}-\db^{(n-1)}}{\triangle t^{(n)}}
    \label{eq:faraday_discrete}
\end{equation}
\begin{equation}
  \dcurl \MfMui \db^{(n)} - \Me \dj^{(n)} = \mathbf{s}_e^{(n)}, \\
  \label{eq:ampere_discrete}
\end{equation}
\begin{equation}
  \Me\dj^{(n)} = \Mes{A}^{(n)}\de^{\ (n)} + \dj_{pol}^{(n-1)}
  \label{eq:ohmslaw_discrete}
\end{equation}
where
\begin{align}
  \dj_{pol}^{(n-1)} = \sum_{k=1}^{n-1} \frac{\triangle t^{(k)}}{2} \Big(\Mes{\dsig (n, k-1)} \e^{\ (k-1)} + \Mes{\dsig (n, k)} \de^{\ (k)} \Big) \nonumber \\
  +  \Mes{\kappa} \de^{\ (n-1)}
\end{align}
Here, $\mathbf{C}$ is the discrete edge-curl operator; $\mathbf{M}^e$ and $\mathbf{M}^f$ are the edge and face inner-product matrices, respectively. For an inner-product matrix, the subscript indicates corresponding physical property (e.g. $M^{f}_{\mu^{-1}}$: the face inner-product matrix for $\mu^{-1}$).

Rearranging the above equations to solve for $\de$ yields:
\begin{align}
  \Big(\dcurl^T \MfMui \dcurl + \frac{1}{\triangle t^{(n)}} \Mes{A}^{(n)}\Big) \de^{(n)} \nonumber \\
  = - \frac{1}{\triangle t^{(n)}} (\mathbf{s}_e^{(n)}-\mathbf{s}_e^{(n-1)})
    + \frac{1}{\triangle t^{(n)}} \Me \dj^{(n-1)} - \frac{1}{\triangle t^{(n)}} \dj_{pol}^{(n-1)}
    \label{eq: discrete_e_solution}
\end{align}
By solving the above equation at each time step, we obtain $\de$. The measured data for AEM are often $-db/dt$ , which can be computed as
\begin{equation}
    \mathbf{db/dt} = -\dcurl \de
    \label{eq: discrete_dbdt_from_e}
\end{equation}
The measured data at a receiver loop can be expressed as
\begin{equation}
    \mathbf{d} = \mathbf{P} (-\mathbf{db/dt})
    \label{eq: discrete_dbdt_data}
\end{equation}
where $\mathbf{P}$ is an interpolation matrix, which projects $\mathbf{db/dt}$ fields, defined in a 3D domain, to a receiver location, and samples those fields at the measured time channels. For discretization of eqs. (~\ref{eq: discrete_e_solution}) to (~\ref{eq: discrete_dbdt_data}) we use, \SimPEG's mesh toolbox. The developed code is open-source as a \textsc{SimPEG-EMIP} package (\url{https://github.com/sgkang/simpegEMIP})

\subsection{Handling the singularity at $\sigma(t=0)$}
\label{app: Handling singularity}
The SE conductivity, $\sigma_{se}(t)$ at $t$=0, is singular, whereas its integral is well-defined, as shown in eq. (\ref{eq: sigma_se_step_off}). When discretizing eq. (\ref{eq:ohmslaw}), this singularity will be problematic. In particular, the issue occurs at the last time segment ($k=n$) of the convolution term in eq. (\ref{eq:ohmslaw_time_original}), which can be written in continuous form:
\begin{equation}
  \int_{t^{n-1}}^{t^{n}} \dsig(u) \e(t-u) du
  \label{eq: convoluation_last_segment}
\end{equation}
This problem also occurs when the Cole-Cole function is used. \cite{marchant2015} (see page 31) tackled this issue by approximating $\e$ at this time segment as a linear function:
\begin{equation}
  \e(t) = \frac{t^{n}-t}{\triangle t^{(n)}} \e^{ \ (n-1)} + \frac{t-t^{n-1}}{\triangle t^{(n)}} \e^{ \ (n-1)}, \ \text{when} \ (t^{(n-1)}\leq t \leq t^{(n)})
\end{equation}
Then by substituting this in to eq. (\ref{eq: convoluation_last_segment}), and evaluating the integration, the discrete form of eq. (\ref{eq: convoluation_last_segment}) is obtained:
\begin{equation}
  \int_{t^{n-1}}^{t^{n}} \dsig(u) \e(t-u) du \simeq
  \kappa(\triangle t^{(n)}) \e^{\ (n-1)} + \gamma(\triangle t^{(n)}) \e^{\ (n)}
\end{equation}

To obtain $\gamma(\triangle t^{(n)})$ and $\kappa(\triangle t^{(n)})$, we use the same trick. Integration of $\dsig(t)$ is not possible, so by Taylor expanding, we obtain an approximate form of $\dsig(t)$ which is valid for small $t$:
\begin{align}
  \dsig (t) = -\siginf \eta_{se}t^{-1}(\frac{t}{\tau_{se}})^{c_{se}}exp\Big(-(\frac{t}{\tau_{se}})^{c_{se}}\Big) \nonumber \\
  \simeq -\siginf \eta_{se}t^{-1}(\frac{t}{\tau_{se}})^{c_{se}}\Big(1-(\frac{t}{\tau_{se}})^{c_{se}}\Big) \nonumber  \\
  = -\siginf \eta_{se}t^{-1}\Big((\frac{t}{\tau_{se}})^{c_{se}}-(\frac{t}{\tau_{se}})^{2c_{se}}\Big)
  \label{eq:se_conductivity_approximate}
\end{align}
By substituting eq. (\ref{eq:se_conductivity_approximate}) into eq. (\ref{eq: convoluation_last_segment}) and evaluating the integral, we finally obtain
\begin{equation}
  \gamma(\triangle t^{(n)}) = \siginf m \Big( \frac{(\triangle t^{(n)})^{c_{se}}}{c_{se}(c_{se}+1)}-
  \frac{(\triangle t^{(n)})^{2c_{se}}}{2c_{se}(2c_{se}+1)\tau_{se}^{c_{se}}} \Big)
\end{equation}
\begin{equation}
  \kappa(\triangle t^{(n)}) = \siginf m \Big( \frac{(\triangle t^{(n)})^{c_{se}}}{c_{se}+1}-
  \frac{(\triangle t^{(n)})^{2c_{se}}}{(2c_{se}+1)\tau_{se}^{c_{se}}} \Big)
\end{equation}

\begin{figure}[htb]
  \centering
  \includegraphics[width=1.0\textwidth]{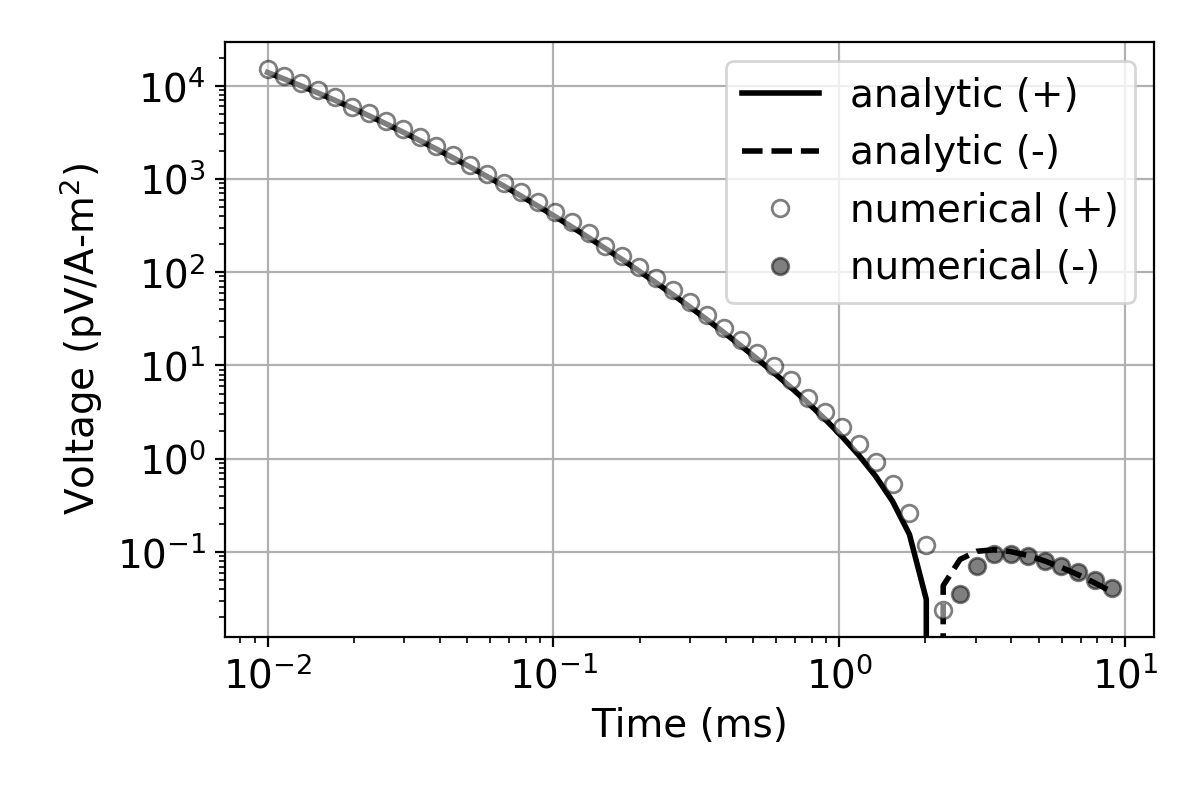}
  \caption{Comparison of numerical and analytic solutions for halfspace earth. SE parameters of the halfspace earth are $\siginf$=0.05S/m, $\eta_{se}$=0.7, $\tau_{se}$=4ms, $c_{se}$=0.6; corresponding Cole-Cole parameters are: $\eta_{cc}$=0.8, $\tau_{cc}$=5ms, $c_{cc}$=0.6. Lines and circles distinguish analytic and numerical solutions.}
  \label{fig:analytic_test}
\end{figure}

\append{Analytic test}
\label{app: Analytic test}
To test the developed \textsc{SimPEG-EMIP} code, we compare our numerical solution with an analytic solution. A halfspace earth is assumed. The conductivity of the halfspace is 0.05 S/m and its SE parameters are: $\eta_{se}$=0.7, $\tau_{se}$=4ms, $c_{se}$=0.6. Corresponding Cole-Cole parameters are: $\eta_{cc}$=0.8, $\tau_{cc}$=0.005s, $c_{cc}$=0.6.
For the spatial discretization, a 2D cylindrically symmetric mesh is used; the smallest cell size is 6.5m $\times$ 5m.
A horizontal source loop is located 30m above the surface. A step-off waveform is used for the input current and a horizontal receiver loop measuring the voltage (equivalent to -$db_z/dt$) is coincident with the source loop. Data are measured in the off-time over the time-range: 10$^{-2}$-10 ms.
Fig. \ref{fig:analytic_test} shows comparison between analytic and numerical solutions; they match well except for a small shift in the time of the zero-crossing, the two solutions are in good agreement.

\clearpage
\bibliographystyle{seg}  
\bibliography{biblio}
\end{document}